\documentclass[pre,twocolumn,showpacs,preprintnumbers,amsmath,amssymb]{revtex4}

\usepackage{graphicx}
\begin{document}


\title{Morphology transitions induced by chemotherapy in carcinomas ``in situ''}

\author{S. C. Ferreira Jr.}
 \email{silviojr@fisica.ufmg.br}

\affiliation{Departamento de F\'{\i}sica, Instituto de Ci\^encias Exatas, Universidade
Federal de Minas Gerais, \\Caixa Postal 702, 30161-970 Belo Horizonte, Minas Gerais,
Brazil}

\author{M. L. Martins}
 \email{mmartins@ufv.br}

\affiliation{Departamento de F\'{\i}sica, Universidade Federal de Vi\c{c}osa, 36571-000,
Vi\c{c}osa, Minas Gerais, Brazil}

\author{M. J. Vilela}
 \email{marcelo@ufv.br}

\affiliation{Departamento de Biologia Animal, Universidade Federal de Vi\c{c}osa,
36571-000, Vi\c{c}osa, Minas Gerais, Brazil}

\date{\today}

\begin{abstract}
Recently, we have proposed a nutrient-limited model for the avascular growth of tumors
including cell proliferation, motility and death \cite{jr}, that, qualitatively
reproduces commonly observed morphologies for carcinomas {\it in situ}. In the present
work, we analyze the effects of distinct chemotherapeutic strategies on the patterns,
scaling and growth laws obtained for such nutrient-limited model. Two kinds of
chemotherapeutic strategies were considered, namely, those that kill cancer cells and
those that block cell mitosis but allows the cell to survive for some time. Depending on
the chemotherapeutic schedule used, the tumors are completely eliminated, reach a
stationary size or grow following power laws. The model suggests that the scaling
properties of the tumors are not affected by the mild cytotoxic treatments, although a
reduction in growth rates and an increase in invasiveness are observed. For the
strategies based on antimitotic drugs a morphological transition in which compact tumors
become more fractal under aggressive treatments was seen.

\end{abstract}


\maketitle

\markright{\small \it S. C. Ferreira Jr., et. al.}

\section{Introduction}

Cancer is the uncontrolled cellular growth in which neoplastic cells invade adjacent
normal tissues and give rise to secondary tumors (metastasis) on tissues or organs
distant from its primary origin \cite{hanahan}. A cancer that remains confined within a
messed surface, without break of the underlying basement membrane, is referred to as
carcinoma ``in situ''. ``In situ'' carcinoma is characterized by intense cytological
atypia, necrosis, and frequent and abnormal mitosis, the tumor cells being arranged in
various distinctive patterns \cite{mcgee}. A malignant tumor is derived from a single
cell that, years before the tumor becomes clinically detectable, began an inappropriate
pathway of proliferation \cite{weinberg}. Although cancers are extremely diverse and
heterogeneous, a small number of pivotal steps associated to both deregulated cell
proliferation and suppressed cell death are required for the development of any and all
tumors. Indeed, all neoplasms evolve accordingly to an universal
scheme~\cite{clark,evan}. In the struggle against cancer, surgical removal, chemotherapy
and/or radiotherapy are the most commonly used treatment for the complete eradication of
the tumor mass. Nowadays, new approaches, such as immunotoxins \cite{old}, gene
\cite{navarro}, anti-angiogenic \cite{folkman} and virus \cite{bischoff} therapies, are
being developed and have been successfully used in the treatment of several kinds of
experimental and human tumors.

Mathematical models are always used as a tentative for describing cancer growth. In
particular, numerous  models based on classical reaction-diffusion equations have been
proposed to investigate the growth of tumor spheroids \cite{pettet}, cancer progress and
its interaction with the immune system \cite{bellomo}, and the tumor angiogenesis
\cite{levine,deangelis}. Fractal growth patterns in gliomas (brain tumors) were recently
investigated by Sander and Deisboeck \cite{sander}. Scalerandi {\it et. al.}, using the
local interaction simulation approach (LISA), formulated models for the evolution of
avascular tumors \cite{scalerandi1} and tumor cords \cite{scalerandi2} under nutrient
competition. They also analyzed the effects of tumor vascularization \cite{scalerandi3}
on cancer growth. Recently, we have studied a diffusion-limited model for the growth of
carcinomas {``in situ''} in which cell proliferation, motility and death are locally
controlled by growth factors \cite{jr1,jr2}. This model generates compact, connected and
disconnected morphologies characterized by Gompertzian growth in time and distinct
scaling laws for the tumor interfaces. In order to generate papillary and ramified
morphologies found in many epithelial cancers and trichoblastomas, we were led to analyze
the effects of nutrient competition in cancer development \cite{jr}.

In addition to the vast literature dedicated to tumor growth modeling, many research
papers addressing cancer therapies have recently been published. Indeed, cancer cell
kinetics under treatments using antimitotic agents \cite{kozusco,montalenti},
radiotherapy \cite{sachs}, virus that replicate selectively in tumor cells \cite{wu},
anti-angiogenic chemicals \cite{scalerandi4}, as wells as the effects of tumor drug
resistance and tumor vasculature on chemotherapies \cite{jackson}, were studied using
mathematical models.

In this paper it is analyzed the effects of distinct chemotherapeutic strategies in the
model for the growth of avascular tumors proposed by us \cite{jr}. Specifically, we are
interested in possible changes in the tumor patterns, scaling and growth laws reported on
the original model triggered by chemotherapies. The paper is organized as follows. In
section \ref{nutmodel} the nutrient-limited model for cancer growth and its main results
are reviewed. In section \ref{chemo}, chemotherapeutic treatments that aim to kill cancer
cells are introduced in the framework of the present model and their effects discussed.
In section \ref{anti}, a model for chemotherapy with antimitotic agents is considered.
Finally, we draw some conclusions in section \ref{conclu}.

\section{The nutrient-limited model}
\label{nutmodel}

Our nutrient-limited model combines macroscopic reaction-diffusion equations, describing
the nutrient field concentration, with microscopic stochastic rules governing the actions
of individual tumor cells. The basic principles included in the model are cell
proliferation, motility and death as well as competition for nutrients among cancer and
normal cells. Also, the nutrient concentration field locally controls cell division, migration and death.

\subsection{The model}
\subsubsection{The tissue}
The tissue, represented by a square lattice of size $(L+1)\times (L+1)$ and lattice
constant $\Delta$, is fed by a single capillary vessel at $x=0$, the top of the lattice.
The capillary is the unique source from which nutrients diffuse through the tissue
towards the individual cells. Although a tumor mass is composed of different cell
subpopulations~\cite{clark}, the model considers only three types: normal, cancer and
tumor dead cells. Any site, with coordinates $\vec{x}=(i\Delta,j\Delta)$,
$i,j=0,1,2,\ldots,L$, is occupied by only one of these cell types. In contrast to the
normal cells, one or more cancer cells can pile up in a given site. In turn, dead tumor
cells are inert. Thus, each lattice site can be thought of as a group of cells in which
the normal, dead and cancer cell populations assume one of the possible values
$\sigma_n(\vec{x})= \sigma_d(\vec{x})=0,1$ and $\sigma_c(\vec{x})=0,1,2,\ldots$,
respectively. Accordingly the theory of the monoclonal origin of cancer~\cite{weinberg},
a single cancer cell at $y=L\Delta/2$ and at a distance $X$ from the capillary is
introduced in the normal tissue. Periodic boundary conditions along the horizontal axis
are used. The row $i=0$ represents the capillary vessel and the sites with $i=L+1$
constitute the external border of the tissue. This geometry is particularly adequate to
describe the growth of carcinomas (epithelial tumors) ``in situ'' because the present
model consider that the tumor mass receive nutrients only by diffusion from the capillary
vessel. However, the model can be extended to others cancers.

\subsubsection{The nutrients}

The nutrients are divided into two classes: essential to cell proliferation, those
necessary to DNA synthesis, and nonessential to cell division. The essential and
nonessential nutrients are described by the concentration fields $N(\vec{x},t)$ and
$M(\vec{x},t)$, respectively. These nutrient fields obey the dimensionless diffusion
equations (see ref. \cite{jr} for the complete variable transformations)
\begin{eqnarray}
\label{eqdifN} \frac{\partial{N}}{\partial{t}}=\nabla^2 N-\alpha^2
N\sigma_n -\lambda_N \alpha^2 N \sigma_c~~~~~~~~~~~
\end{eqnarray}
and
\begin{eqnarray}
\label{eqdifM} \frac{\partial{M}}{\partial{t}}=\nabla^2 M-\alpha^2
M \sigma_n -\lambda_M \alpha^2 M\sigma_c~~~~~~~~~~~
\end{eqnarray}
in which differentiated nutrient consumption rates for normal and cancer cells by factors
$\lambda_N$ and $\lambda_M$ are assumed. It is important to notice that the model admits
the simplest form for nutrient diffusion, i. e., linear equations with constant
coefficients. Also, $\lambda_N > \lambda_M$ is used, reflecting the larger affinity of
cancer cells for essential nutrients.

The boundary conditions  satisfied by the normalized nutrient concentration fields are
$N(x=0)=M(x=0)=1$, representing the continuous and fixed supply of nutrients provided by
the capillary vessel; $N(y=0)=N(y=L)$ and $M(y=0)=M(y=L)$, corresponding to the periodic
boundary conditions along the y-axis; at last, Neumann boundary conditions,
$\partial{N(x=L)}/\partial{x}=\partial{M(x=L)}/\partial{x}=0$, are imposed to the border
of the tissue.

\subsubsection{Cell dynamics}
\label{dynamics}

In our original model, each tumor cell, randomly selected with equal probability, can
carry out one of three actions: division, migration or death. However, in the present
work we consider just the accommodation that happens during cell mitosis among the
daughter cells. Consequently, each tumor cell can carry out one of two actions.
\begin{enumerate}
  \item {\it Division}. Cancer cells divide by mitosis with probability $P_{div}$. If the
chosen cell is inside the tumor, its daughter will pile up at that site, and
$\sigma_c(\vec{x}) \rightarrow \sigma_c(\vec{x})+1$. Otherwise, if the selected cell is
on the tumor border, its daughter cell will occupy at random one of their nearest
neighbor sites $\vec{x^\prime}$ containing a normal or a necrotic cell and, therefore,
$\sigma_c(\vec{x^\prime})=1$ and $\sigma_{n,d}(\vec{x^\prime})=0$. The mitotic
probability $P_{div}$ is determined by the concentration per cancer cell of the essential
nutrients $N$ present on the microenvironment of the selected cell:
  \begin{equation}
    \label{pdiv}
    P_{div}(\vec{x})=1-\exp\left[- \left( \frac{N(\vec{x})}{\sigma_c(\vec{x}) \; \theta_{div}} \right)^2
    \right].
  \end{equation}
  The Gaussian term is included in order to produce a sigmoid curve saturated to the
  unity, and the model parameter $\theta_{div}$ controls the shape of this sigmoid.

  \item {\it Death}. Cancer cells die with probability
   $P_{del}$. Thus, $\sigma_c(\vec{x}) \rightarrow \sigma_c(\vec{x})-1$ and
   $\sigma_d(\vec{x})=1$ when $\sigma_c$ vanishes. The cell death probability $P_{del}$ is
   determined by the concentration per cancer cells of the nonessential nutrients $M$
   present on the microenvironment of the selected cell:
  \begin{equation}
    \label{pdel}
   P_{del}(\vec{x})=\exp\left[-\left( \frac{M(\vec{x})}{\sigma_c(\vec{x}) \; \theta_{del}} \right)^2 \right],
  \end{equation}
  i. e., a Gaussian distribution whose variance depends on the model parameter $\theta_{del}$.
\end{enumerate}

The biological basis for these cell dynamic rules can be found in reference \cite{jr}.
However, it is worthwhile to notice that from the point of view of the so-called kinetic
cellular theory, which provides a general framework for the statistical description of
the population dynamics of interacting cells \cite{bellomo}, the local probabilities
$P_{div}$ and $P_{del}$ can be thought as an effective kinetic cellular model.

\subsubsection{Computer implementation}

The growth model simulations were implemented using the following procedure. At each time
step $T$, Eqs. (\ref{eqdifN}) and (\ref{eqdifM}) are numerically solved in the stationary
state (${\partial N}/{\partial t}={\partial M}/{\partial t}=0$) through relaxation
methods. Provided the nutrient concentration at any lattice site, $N_{C}(T)$ cancer cells
are sequentially selected at random with equal probability. For each one of them, a
tentative action (division or death) is randomly chosen with equal probability and the
time is incremented by $\Delta T=1/N_{C}(T)$. The selected cell action will be
implemented or not according to the correspondent local probabilities determined by Eqs.
(\ref{pdiv}) or (\ref{pdel}). At the end of this sequence of $N_C(T)$ tentatives, a new
time step begins and the entire procedure (solution of the diffusion equations and
application of the cell dynamics) is iterated. The simulations stop if any tumor cell
reaches the capillary vessel.

\subsection{Main results}

The model reproduces commonly observed tumor morphologies including  the papillary,
compact and ramified patterns shown in Fig.  \ref{padterapia}. The nutrient consumption
by normal and cancer cells, controlled by the model parameters $\alpha$, $\lambda_N$ and
$\lambda_M$, plays a central role in morphology selection. For small values of these
parameters, corresponding to growth conditions in which individual cells demand small
nutrient supplies, the patterns tend to be compact and circular (Fig. \ref{padterapia}a).
However, if the mitotic rate of cancer cells is small due to the large amount of
nutrients demanded for cell division, generating a significant competition for nutrients,
these compact patterns progressively assume papillary-like morphologies (Fig.
\ref{padterapia}b). At high nutrient consumption rates these papillary patterns become
the rule and, for low cancer cell division, continuously transform into thin tips,
filaments or chords of cells (Fig.  \ref{padterapia}c). Also, the model generates
patterns with a inner core of died cells for high nutrient consumption or cell division
rates (Fig. \ref{padterapia}d). As observed in {``in vivo''} tumors and {``in vitro''}
multicell spheroids~\cite{pettet}, these simulated patterns consist of three distinct
regions: a central necrotic core, an inner rim of quiescent cancer cells and a narrow
outer shell of proliferating cells.

The tumor patterns generated by our nutrient-limited model were characterized by its
gyration radius $R_g$, total number of cancer cells $N_C$, and number of sites on tumor
periphery $S$ (including the surface of holes, if any). The gyration radius $R_g$ is
defined as
\begin{equation}
R_g=\left(  \frac{1}{n} \sum_{i=1}^{n} r_{i}^2 \right)^{1/2}
\end{equation}
where $n$ is the number of sites occupied by the pattern (cancer or necrotic cells) and
$r_i$ is the distance of the occupied site $i$ from the tumor mass center. These
quantities could be related to clinically important criteria such as progress curves,
growth rates (volumetric doubling time) at given radii, proliferative and necrotic
fractions of the tumor.

The progress in time of cancer cell populations for all the simulated patterns follows
Gompertz curves,
\begin{equation}
N_C(T)=A \exp \left[-\exp(-k(T-T_c))\right],
\end{equation}
representing an initial exponential growth that is saturated in larger times.

Alternatively, as a function of the number of sites occupied by the pattern $n$, both
$R_g$ and $S$ obey power law scalings given by $R_g \sim n^{\nu}$ and $S \sim
n^{\sigma}$, respectively. For solid patterns these exponents are $\nu \sim 0.5$ and
$\sigma \sim 0.5$, corresponding to effective circular and non fractal patterns. As the
nutrient consumption increases, the patterns tend to papillary-like shapes for which the
exponent $\sigma$ increases towards the value $1$ and the exponent $\nu$ varies in the
range $[0.50,0.60]$, indicating a fractal morphology for the tumor.

\section{Cytotoxic chemotherapeutic drugs}
\label{chemo}

The primary aim of the antitumoral treatment with chemotherapeutic drugs is to kill or at
least to stop the proliferation of the cancer cells. In general, the drugs should act
only on proliferating cells, mainly the cancer ones. However, drugs also destroy
proliferating normal cells promoting several collateral effects \cite{hellman}. Indeed,
epithelial cells from the respiratory and gastro-intestinal systems, which frequently
reproduce in order to substitute their dead counterparts, are strongly affected. In this
section, we analyze a simple chemotherapeutic model in which the complex details of the
cell-cycling responses to the drugs are taking into account by an effective kinetic
model.

\subsection{The model}

\begin{figure}[hbt]
\begin{center}
\resizebox{8.5cm}{!}{\includegraphics{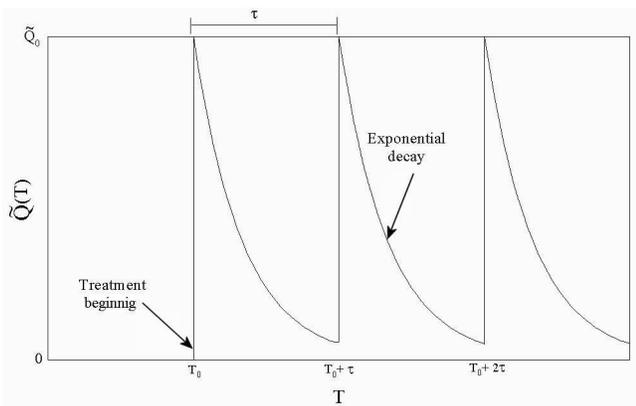}}
\end{center}
\vspace{-0.5cm} \caption{\small Temporal profile of drugs concentration at the capillary
vessel.} \label{dose}
\end{figure}

As used in previous works \cite{panetta2}, the chemotherapy is modelled by a periodic
delivery of cytotoxic drugs through the same capillary vessel supplying the nutrients to
the tissue. Several cytotoxic drugs and their properties were exhaustively studied such
as amsacrine, cisplatin, cyclophosphamide, cytarabine, mustine, and anthracycline
\cite{oxford}. Here, the numerous barriers involved in tumor drug delivery \cite{jain}
were not considered, and the treatment begins when the tumor mass contains $N_0$ cancer
cells. When a dose is applied, the drug concentration in the capillary assumes a value
$\widetilde{Q}_0$. By hypothesis, this concentration level decays exponentially in time,
simulating the gradual drug elimination by the organism. New doses are periodically
administered at time intervals $\tau$. So, the drug concentration at the capillary,
$\widetilde{Q}(T)$, is a function of time written as
\begin{equation}
\label{eqdose}
  \widetilde{Q}(T)=\left \{ \begin{array}{l}
              0,\mbox{~~if~~} T<T_0 \\
              \widetilde{Q}_0 \exp{\lbrack-(T-l\tau)/T_\times\rbrack},\mbox{~~if~~} T_0+l\tau\leq T<\\
              ~~~~~~~~~~~~~~~~~~~~~~~~~~~~~~~~~~~T_0+(l+1)\tau
             \end{array} \right.,
\end{equation}
where $l=0,1,2\ldots$ is the number of applied doses, $T_\times$ is the characteristic
time for the drug elimination by the organism and $T_0$ is the time at which the
treatment begins. The functional form of $\widetilde{Q}(T)$ is shown in Fig.  \ref{dose}.

As the nutrients, drugs diffuse from the capillary vessel toward the individual cells and
their concentration $Q(\vec{x},t)$ at any lattice site at each time step is given by the
stationary solution of the diffusion equation
\begin{equation}
 \frac{\partial{Q}}{\partial{t}}=D_Q\nabla^2
Q-\Gamma^2 Q-\lambda_Q \Gamma^2 Q \sigma_c.
\label{eqdifQ}
\end{equation}
The first, second and third terms on the right hand side represent drug diffusion,
natural degradation and absorption by cancer cells, respectively. Again, this equation is
the simplest one describing the diffusion phenomenon. Eq. (\ref{eqdifQ}) is written using
the same dimensionless variables of (\ref{eqdifN}) and (\ref{eqdifM}). Also, the
parameters $\widetilde{Q}_0$ and $D_Q$ can be make equal to the unity without generality
loss. The boundary conditions are the same used for the nutrient fields, except at the
capillary vessel where the concentration at each time step is given by Eq.
(\ref{eqdose}).

Finally, a single change is introduced into the cell dynamics described in section
\ref{nutmodel}: an additional chance of death occurs whenever a cancer cell enters
mitosis. Thus, every time a cancer cell divides, each one of the generated cells can die
with probability
\begin{equation}
  P_{del}^{(Q)}(\vec{x})=1-\exp\left[-\left( \frac{Q(\vec{x})}{\sigma_c(\vec{x}) \; \theta_{del}^{(Q)}} \right)^2
  \right].
  \label{pdelQ}
\end{equation}
The parameter $\theta^{(Q)}_{del}$ controls the cell sensitivity to the drugs. Also,
since in the original nutrient-limited model the normal cells do not take part on the
cell dynamics, i. e., they do not divide or die, we disregard the chemotherapeutic
effects on normal cells.

\subsection{Results}

The main aim of this work was to investigate the effects of treatments on the various
morphologies, scaling and growth laws observed in the original model. Thus, for each
morphology, determined by the fixed parameters reported in table \ref{tabpar}, the
treatment parameters, namely, $\tau$ (the dose period), $\theta^{(Q)}_{del}$ (drug
efficiency) and $N_0$ (tumor size at the treatment beginning) were varied. Such
parameters can be directly tested in the laboratory. The remaining parameters
($\lambda_Q$, $\Gamma$ and $T_\times$) associated to drug diffusion and elimination were
also fixed in all the simulations.

\begin{table}[hbt]
\begin{center}
\begin{tabular}{cccccccccc}\hline\hline
Morphology&$\lambda_N$&$\lambda_M$&$\lambda_Q$&$\alpha$&$\Gamma$&$\theta_{div}$& $\theta_{del}$&$T_\times$\\\hline
Compact  &    $25$   &   $10$    &    $10$   & $2/L$  & $2/L$  &     $0.3$    &  $0.03$     &   $4$\\
Papillary   &    $200$  &   $10$    &    $10$   & $2/L$  & $2/L$  &     $0.3$ &   $0.03$     &   $4$\\
Ramified&    $200$  &   $10$    &    $10$   & $3/L$  & $2/L$  &     $0.3$    &    $0.01$     &   $4$\\
Necrotic&   $50$   &   $25$    &    $10$   & $3/L$  & $2/L$  &     $0.3$    & $0.03$     &   $4$\\
\hline\hline
\end{tabular}
\end{center}
\vspace{0.0cm} \caption{Fixed parameters used in tumor growth simulations under
chemotherapeutic treatment for each morphology type.} \label{tabpar}
\end{table}

\begin{figure}[hbt]
\begin{center}
\resizebox{6.0cm}{!}{\includegraphics{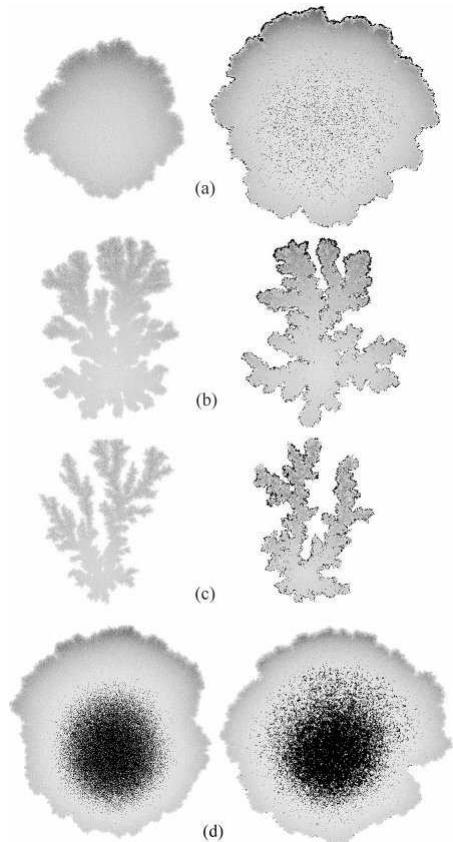}}
\end{center}
\caption{Tumor growth patterns generated by the limited-nutrient model under
chemotherapeutic treatment. The patterns are drawn in a gray scale where the darker gray
regions represent higher cancer cell populations, the black points represent the sites
occupied by necrotic cells, and the white regions represent the normal tissue. The tissue
size is $500\times500$, and the initial ``cancer seed'' is 300 sites distant from the
capillary. The total number of cancer cells depends on tumor morphology and reach up to
$2\times 10^5$ for compact patterns. Two typical patterns, without (left) and with soft
treatment (right) are shown, for  (a) compact, (b) papillary, (c) ramified morphologies
and (d) patterns with a necrotic core. The fixed parameters used for generate these
morphologies are listed in table \ref{tabpar}. Mild treatment means that the period
between two doses is large ($\tau=20$) and, therefore, the tumor grows continuously. The
other parameters are $\theta_{del}^{(Q)}=0.1$ and $N_0=10^4$.} \label{padterapia}
\end{figure}

In Fig.  \ref{padterapia}, the corresponding patterns for compact, papillary, and
ramified morphologies with and without treatment are shown. In this figure the treatment
is not able to cease tumor development. As can be seen, the morphological tumor patterns
do not change under mild treatment. However, the  regions occupied by the tumors are
larger than those without treatment. These results suggest that the direct attack to the
tumor might be an inadequate treatment strategy. Indeed, the more invasive is the tumor,
higher is the possibility of the cancer cells to reach the capillary vessel and,
consequently, metastasize successfully. Such result is in agreement with the Israel's
claim \cite{lucien} that cancer cells trigger adaptation mechanisms in stress
circumstances similar to those observed in bacterial colonies. Moreover, in our model
neither genetic or epigenetic changes are necessary to explain the increase of tumor
aggressiveness. It naturally emerges from the growth rules. A tumor submitted to
successive chemotherapeutic treatments which do not lead to its complete eradication may
progressively become more resistant, aggressive and malignant.

Very similar scaling laws $R_g\sim n^\nu$ and $S\sim n^\sigma$ for the treated tumors
were observed ($\nu$ and $\sigma$ values are reported in \cite{jr}). The small
differences in the exponents values for the number of tumor peripheral sites vanishes at
the asymptotical limit of tumor size. This exponents invariance suggests that the fractal
morphology is a robust aspect of these tumors and cannot be changed by small
perturbations in the cell microenvironment. Obviously, the scaling laws for $R_g$ and $S$
are not defined for the tumors that cease their growth.


\begin{figure}[hbt]
\begin{center}
\resizebox{8.5cm}{!}{\includegraphics{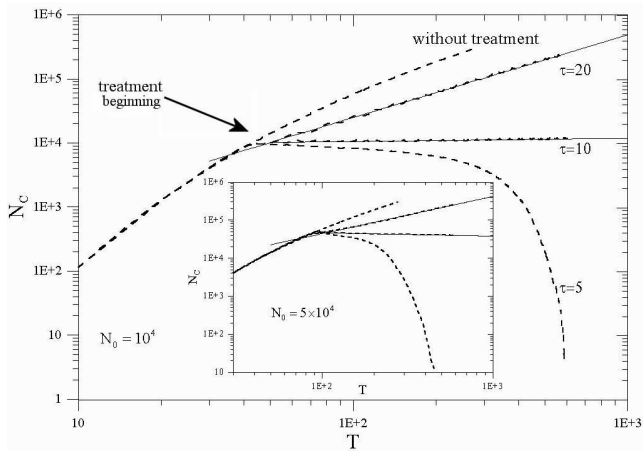}}
\end{center}
\vspace{-0.5cm} \caption{\small Growth curves for compact patterns (dashed lines). Three
dose intervals were tested ($\tau=5,10,20$) in tumors with two distinct initial sizes:
$N_0=10^4$ and $N_0=5\times 10^4$ (inset). The exponents of the power laws (slopes of the
straight lines) are  1.3 (0.98) for the tumors with $N_0=10^4$ ($N_0=5\times 10^4$) when
$\tau=20$, and 0.05 (-0.09) when $\tau=10$. } \label{ncterapia}
\end{figure}

We have also studied the influence of the treatment parameters ($\tau$,
$\theta^{(Q)}_{del}$, $N_0$) on the tumor growth curves. In Fig. \ref{ncterapia} the
evolution in time of the total number of cancer cells for compact tumors is shown.
Depending on dose period $\tau$ the tumors may disappear, saturate their growth or
progress according to power laws. Actually, the time interval between two consecutive
doses is a fundamental clinical feature determining the treatment success. The power law
regime observed when $\tau$ is large means a slow progress, neatly contrasting with the
exponential growth present in the Gompertz law describing tumor progress without
chemotherapy. The tumor size at the beginning of the treatment is another important
factor. Indeed, the exponents of the power laws characterizing the tumor growth are
smaller when the initial tumor size is bigger. This scenario seems to be universal.
Sometimes, as shown Fig. \ref{ncterapia}, compact tumors which began to receive drug
doses at regular intervals of $\tau=5$ when it contained $N_0=10^4$ cancer cells were
eliminated slower than another one that began to be treated with $N_0=5\times 10^4$
cancer cells. However, this is not the rule. For example, simulations of papillary tumors
indicate that smaller tumors are faster eliminated when $\tau=5$. Finally, since the
growth law exponents depend on the parameter sets used, they are not universal.
Concerning the $\theta^{(Q)}_{del}$ parameter, it just modifies the treatment efficiency.

\begin{figure}[hbt]
\begin{center}
\resizebox{8.5cm}{!}{\includegraphics{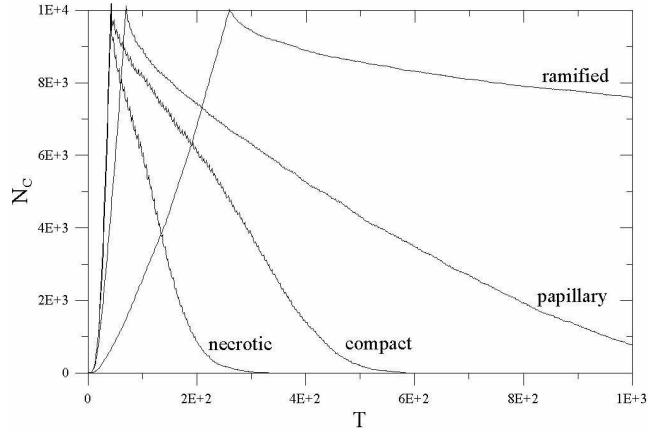}}
\end{center}
\vspace{-0.5cm} \caption{\small Growth curves for distinct tumor morphologies. The doses
were applied at intervals $\tau=5$ in tumors with initial sizes $N_0=10^4$.} \label{cpr}
\end{figure}

Fractal tumors are more resistant to treatments. In Fig. \ref{cpr}, the growth curves
$N_C\times T$ are drawn for compact, papillary, ramified and necrotic morphologies shown
in Fig.  \ref{padterapia}. One can see that the more fractal is the tumor larger is the
required time to eliminate it. Indeed, the lower growth rates of fractal tumors imply in
a large fraction of cancer cells maintained at a quiescent state. So, since
chemotherapeutic drugs considered in this model only act in the dividing cells, the
treatment becomes inefficient when the major fraction of the cancer cells are quiescent.

In addition, we have studied tumor patterns exhibiting a central necrotic core. The
results for cancer growth and the correspondent power laws are similar to those found for
compact patterns. The density of cancer cells and their average division rates through
the growth patterns are not significatively altered when the tumors are submitted to mild
treatments (long time intervals between consecutive drug doses). Moreover, their growth
patterns are very similar to those found in the untreated counterparts. Finally, tumors
exhibiting necrotic cores are more easily eliminated when shorter periods of the drug
administration are considered (Fig. \ref{cpr}).

\section{Antimitotic chemotherapeutic drugs}
\label{anti}

\begin{figure}[hbt]
\begin{center}
\resizebox{8.0cm}{!}{\includegraphics{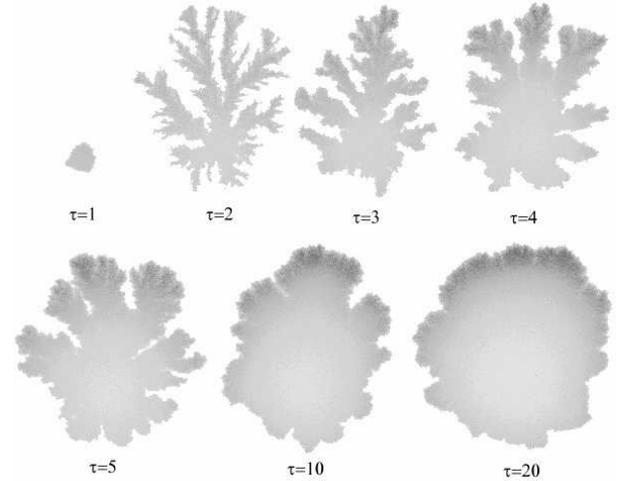}}
\end{center}
\vspace{-0.5cm} \caption{\small Compact patterns under treatment with antimitotic drugs.
The dose period $\tau$ was varied and $\theta_{div}^{(Q)}=0.3$ was fixed. The other
parameters are indicated in table \ref{tabpar}. Tumors become more and more fractal as
the dose period decreases. At a critical period the tumors reach a frozen state in which
their size remain constant during all the treatment.} \label{antipadt}
\end{figure}

Several drugs used in cancer chemotherapies do not kill cancer cells. Instead, they aim
to stop the cell cycle in specific checkpoints. As a consequence of that, the tumors
cease or slow down their growth rate. Examples of such drugs include the antimitotic
agent Curacin A that blocks the cell cycle in the M-phase, mitomycin C, doxorubicin, and
aclarubicin, among others \cite{oxford}. In order to analyze the effects of drugs that
inhibit cell division on tumor patterns, we introduce a very simple change in the cell
dynamics of the model described in the previous section. Instead of an additional death
probability given by Eq. (\ref{pdelQ}), the division probability is modified by including
an additional term in Eq. (\ref{pdiv}). So, the new division probability becomes
\begin{equation}
  \label{pdivQ}
  P_{div}=1-\exp\left[- \left( \frac{N}{\sigma_c\theta_{div}} \right)^2
  +\left( \frac{Q}{\sigma_c\theta_{div}^{(Q)}} \right)^2 \right],
\end{equation}
in which the first and second terms in the exponential argument compete between
themselves stimulating and inhibiting the cell mitosis, respectively. The parameter
$\theta_{div}^{(Q)}$ controls the drug influence on the cell division. Obviously, if the
exponential argument is larger than the unity then $P_{div}\equiv 0$.

\begin{figure}[hbt]
\begin{center}
\resizebox{8.5cm}{!}{\includegraphics{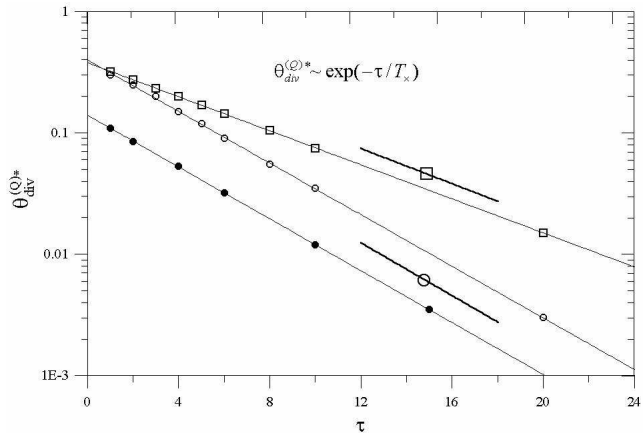}}
\end{center}
\vspace{-0.5cm} \caption{\small Critical values $\theta_{div}^{(Q)*}$ as a function of
the dose period $\tau$. The symbols represent simulational data and the straight lines
the respective exponential fittings. The open circles refer to the simulations using the
parameters for the compact patterns listed in table \ref{tabpar}. The squares (filled
circles) represent the same simulations except that $T_\times=6$ ($\theta_{div}=0.1$).
Also, the slopes $1/4$ (circle) and $1/6$ (square) are indicated. In order to estimate
$\theta_{div}^{(Q)*}$ we realized $20$ independent simulations for each value of
$\theta_{div}^{(Q)}$  and considered that the growth failure for all these tentatives
means tumor latency. } \label{ndivqt}
\end{figure}

The biological interpretation of the present model is very distinct from that considered
in the last section. Now, the drugs modify the intrinsic characteristics of the cancer
cells and, consequently, the tumors must behave differently from those treated with
cytotoxic agents. Indeed, as one can see in Fig. \ref{antipadt},  a morphological
transition for the tumor patterns when the dose periods decrease is observed . If the
period is sufficiently short the tumor size remains constant during the therapy.
Obviously, the critical period depend on other model parameters, specially on
$\theta_{div}^{(Q)}$. In order to see this dependence, was evaluated the critical value
of $\theta_{div}^{(Q)*}$ for which the tumors cease their growth with probability $1$ as
a function of $\tau$. A mean field analysis of Eq. (\ref{pdivQ}) provides
$\theta_{div}^{(Q)}\sim \exp(-\tau/T_{\times})$, in agreement with the simulations as
indicated in Fig. \ref{ndivqt}. Thus, it was found an exponential decay with an universal
characteristic length $T_\times$ that is independent of the other model parameters. This
law, relating two important clinical parameters, is valid for an wide set of parameter
values. In addition to the compact tumors shown in figure \ref{antipadt}, it was also
simulated papillary tumors under antimitotic treatment, and a similar morphology
transition was observed.

\begin{figure}[hbt]
\begin{center}
\resizebox{8.5cm}{!}{\includegraphics{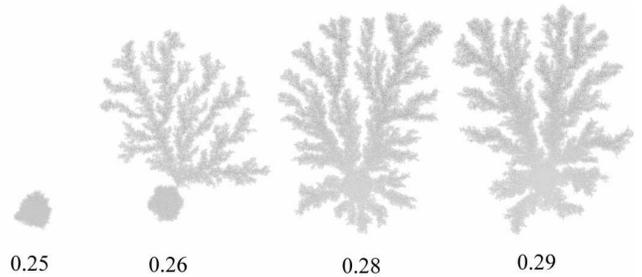}}
\end{center}
\vspace{-0.5cm} \caption{\small Morphological transitions at $\theta_{div}^{(Q)*}$. The
patterns were originally compact and the model parameters are listed in table
\ref{tabpar}. The dose period was fixed at $\tau=2$ and $\theta_{div}^{(Q)}$ varied
around its critical value.} \label{ndivcrt}
\end{figure}

Fig. \ref{ndivcrt} shows that these morphological transitions occurred at well defined
$\theta_{div}^{(Q)}$ values. Near the critical value, instabilities in the
reaction-diffusion equations lead to a branching growth \cite{sander}. Below this
threshold the cell death rate equals cell division, and the tumor growth effectively
stops. This qualitative behavior near the transition will probably not be observed in
experimental tumors, since a very reliable determination of $\theta_{div}^{(Q)}$ is very
difficult. On the other hand, experiments exhibiting both regimes, above and below the
critical value, are realizable today, due to the great advances in drug delivery control.
Such assays might corroborate the complex behaviors predicted  by our model.

\begin{figure}[hbt]
\begin{center}
\resizebox{8.5cm}{!}{\includegraphics{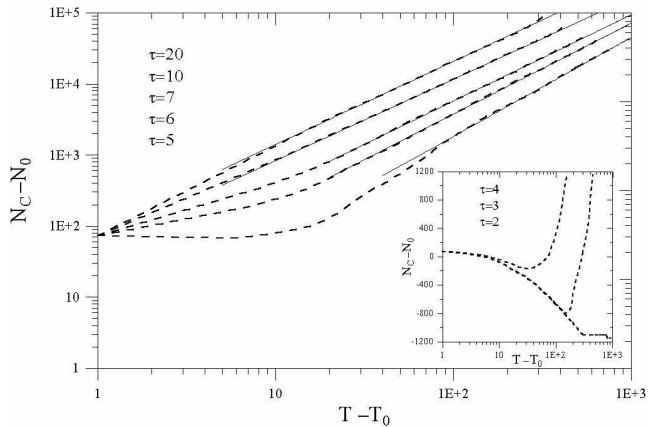}}
\end{center}
\vspace{-0.5cm} \caption{\small Growth curves (dashed lines) for the papillary patterns
under antimitotic therapy. The dose intervals used were $\tau=20,10,7,6,5,4,3,2$ and the
upper curves represent the larger intervals. For the larger dose intervals ($\tau \geq
5$) the tumor growth follows power laws with exponents in the range [1.2,1.4]. For
shorter intervals ($\tau \leq 4$), the curves are shown in a semi-log plot (inset). In
these simulations $\theta_{div}^{(Q)}=0.3$ was used in addition to the parameters of
table \ref{tabpar}.} \label{ncanti}
\end{figure}

The antimitotic treatment affects both the growth and scaling laws. The compact patterns
in the original model, for which the gyration radius and number of occupied sites on the
tumor border scale as the square root of the total number of occupied sites (i. e.
$\nu=\sigma=1/2$), become progressively more fractal as the dose intervals are decreased.
Consequently, it was obtained $\nu>1/2$ and $\sigma>1/2$ indicating fractal tumors with a
fractal dimension given by $d_f=1/\nu$. In general, all the tumors become more fractal
when submitted to antimitotic treatments. So, there is a neat contrast with the
invariance of the fractal dimension of the growing tumors under the cytotoxic treatment
described in the previous section. In Fig. \ref{ncanti} the increase in the number of
cancer cells relative to the treatment beginning ($N_C-N_0$) is plotted, for distinct
dose intervals, as a function of the time afterwards the treatment beginning ($T-T_0$).
These curves suggest power laws for the time evolution of the number of cancer cells with
a weak dependence on the $\tau$ parameter. Indeed, we have $(N_C-N_0) \sim (T-T_0)^\beta$
whit $\beta \in [1.2,1.4]$. As $\tau$ decreases, the number of tumor cells initially
decays for a while, but subsequently recovers its growth (inset of Fig. \ref{ncanti}).
Below a given dose interval, $N_C$ decreases monotonically up to a constant value.

\begin{figure*}[hbt]
\begin{center}
\resizebox{17cm}{!}{\includegraphics{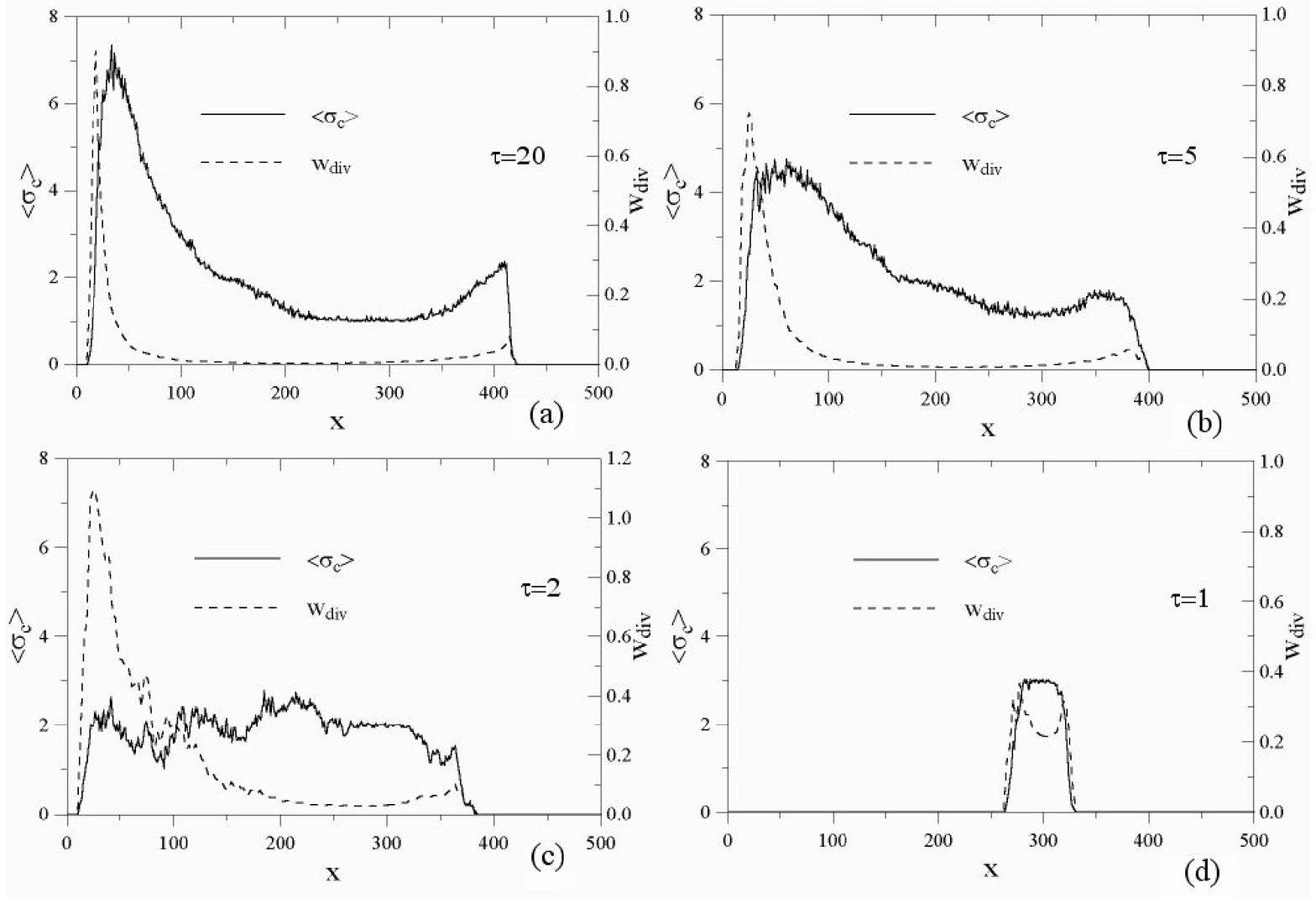}}
\end{center}
\vspace{-0.5cm} \caption{\small Density of cancer cells $\sigma_c$ (continuous lines) and
division rates $w_{div}$ (dashed lines) as a function of the distance from the capillary
vessel along a longitudinal cut of the tumor. The fixed model parameters are those
referent to the compact morphology indicated in table \ref{tabpar} and
$\theta_{div}^{(Q)}=0.3$. The plots correspond to (a) $\tau=20$, (b) $\tau=5$, (c)
$\tau=2$ and (d) $\tau=1$. The left and right vertical axes represent $\sigma_c$ and
$w_{div}$, respectively.} \label{r_t}
\end{figure*}

In order to analyze the cell division rates through the tumor, we computed the average
cancer cell density ($\langle \sigma_c\rangle$) and cell division rate ($w_{div}$) along
a longitudinal cut across the growth pattern. These average values were plotted as a
function of the distance from the capillary vessel. In Fig. \ref{r_t} these plots  are
shown  for compact patterns treated with distinct $\tau$ values. For larger $\tau$ values
(Fig. \ref{r_t}(a)) both, division rate and cancer cell density, have sharp maxima on the
tumor border in front and opposite to the capillary vessel. Notice that the peaks for
division rate are significatively narrower than those for cancer cell density,
demonstrating that the proliferative fraction of the tumor comprises just a small rim
located at the tumor border. As the treatment is intensified (shorter $\tau$ values are
considered) the cell densities through the tumors become more uniform, whereas the
division rates continue exhibiting sharp maxima at the tumor borders (Figs. \ref{r_t}(b)
and (c)). If $\tau$ is sufficiently short in order to halt the tumor growth, the division
rates and cancer densities become uniform along the tumor (Fig. \ref{r_t}(d)). So, in
these cases the division rates are counterbalanced by the death rates, leading to a
vanishing net rate of cancer growth. All these results show that the tumor patterns are
altered when the mitotic properties  of the cancer cells are modified by external agents.

\section{Conclusions}
\label{conclu}

In the present work, a reaction-diffusion model to simulate the effects of chemotherapies
on the growth of carcinomas {``in situ''} have been studied. The model includes cell
death and division, competition among cancer and normal cells by nutrients and periodical
drug administration. Two kinds of chemotherapies, using cytotoxic and antimitotic drugs,
were modelled.

In the cytotoxic model the tumors can be completely eradicated, cease their growth or
grow continuously. The last case occurs when the treatment is inefficient or the
intervals between consecutive doses of drugs are large. Moreover, the morphologies and
scaling laws of the growing tumor patterns are preserved. In contrast, for therapeutic
approaches using antimitotic agents a morphological transition in the tumor patterns was
observed. The growth patterns become progressively more fractal as more effective
treatments (shorter intervals between consecutive doses and/or more efficient drugs) are
considered.

Such morphological transitions are similar to recent studies claiming that bacterial
colonies exposed to non-lethal concentrations of antibiotics exhibit drastic changes in
their growth patterns \cite{ben-jacob}. For bacteria, these changes were imputed to
variations in bacterial properties such as metabolic load and chemotaxis. In turn, normal
and cancer cells cultured in monolayer and collagen gel exhibit a dynamical transition in
their aggregation regimes as an adaptive response to the growth constraints imposed by
high cell population density or long permanence in culture \cite{mendes}. Again, the
results obtained in the present work are in agreement with the point of view that cancer
cells can develop an integrated defense program against stress situations similar to the
response of bacterial colonies facing severe and sustained threats \cite{lucien}.
However, as far as we are concerned, there are no reports on cancer literature concerning
transitions morphologies in the histological patterns of tumors submitted to
chemotherapy. Thus, formal models like the one proposed in this paper, generally not
familiar for the most of the biomedical researchers \cite{complexity}, can guide and
refine new experiments intended to analyze such morphological transitions. Currently,
experiments ``in vitro'' with cancer cells are being performed in our laboratories in
order to investigated those morphological transitions.

It is important to mention that, usually, real effective treatments employ several
therapeutic methods simultaneously \cite{hellman}. So, a more realistic chemotherapeutic
model should consider combined cytotoxic and antimitotic treatments. Nevertheless, in
these cases relevant features such as the robustness of the tumors fractal scaling and
the morphology transitions can be masked. At last, we are modelling others cancer
therapeutic strategies, such as virus and immunotoxins therapies, as well as the
combination of distinct treatments by using reaction-diffusion models like that
considered in this paper.

\begin{acknowledgments}
S.C. Ferreira Jr. would like to thank the UFV Physics Department (Vi\c{c}osa, Brazil) for
the welcoming hospitality and the Professor J. G. Moreira (Departamento de  F\'{\i}sica,
UFMG, Brazil) for fruitful discussions. This work was partially supported by the CNPq an
FAPEMIG -Brazilian agencies.
\end{acknowledgments}


\end{document}